# Three findings to model a quantum-gravitational theory

A Alfonso-Faus

E.U.I.T. Aeronáutica, Plaza Cardenal Cisneros s/n, 28040 Madrid, SPAIN

E- mail: aalfonsofaus@yahoo.es

**Abstract.** In 1967 Zel'dovich expressed the cosmological constant Λ in terms of G, m and ℏ, the gravitational constant, the mass of a fundamental particle and Planck's constant. In 1972 Weinberg expressed m in terms of ℏ, G, the speed of light c and the Hubble parameter H. We proved that both expressions are identical. We also found proportionality between c and H. The critical mass balancing the outward quantum mechanical spreading of the wave function, and its inward gravitational collapse, has been recently estimated. We identify this mass with Zel'dovich's and Weinberg's mass. A semi classical gravity model is reinforced and provides an insight for the modelling of a quantum-gravitational theory. The time evolution of the peak probability density for a free particle, a wave function initially filling the whole Universe, explains the later geometrical properties of the fundamental particles. We prove that they end up acquiring a constant size given by their Compton wavelength. The size of the fundamental particles, as well as their mass, is explained. The three findings converge: Newton's laws of motion and gravitation, while explaining Zel'dovich and Weinberg's relations, also define the mass and size of the fundamental particles when using the Schrödinger-Newton equation.

## 1. Introduction: waves, particles and fields.

Classical interactions imply an interchange of momentum and energy between two or more interacting systems. Gravitation, as Newton thought, implied an interaction at a distance produced by a force. His ideas on gravitational and inertial masses interpreted the gravitational action as an interchange of momentum, a force as treated in his second law of motion. Einstein's special relativity included Newton's laws of motion in the limit of low speeds compared with the speed of light. General Relativity, that includes the gravitational field into the relativity concepts, does not deal with forces. It deals with space-time distorted by masses as equivalent to the presence of a gravitational field. And the masses follow the geometrical trajectories, the geodesics, in accordance with the distorted space-time by the mass sources.

We have quantum mechanics that appears to be a correct interpretation of Nature: it integrates the wave and particle points of view that so much troubled physics when they were considered separately. We have then three ingredients now: waves, particles and fields. A wave-to-particle interaction technique has been used, [1] and [2], proving that all gravitational radii defined by masses inside a proper volume, including the Universe, are true constants of Nature. We also proved, [2], that the speed of light varies proportionally to the Hubble parameter, i.e., inversely proportional to cosmological time t. Then the size of the Universe, as given by the product ct, is a universal constant. It is of the order of its constant gravitational radius.

Gravitational effects are usually referred to as due to the presence of a gravitational field. And the concept of a gravitational field should not only explain the wave properties but also the particle properties of what we understand by gravity. If one believes this we should have a picture of what we mean by a gravitational wave, acting by means of "gravitons" predicted by general relativity. And we should also have a picture of what we mean by a particle of gravitation, a quantum of gravity. There remains the concept of a gravitational force. This has to have the plausible dual interpretation of an interchange of momentum (particles) and a wave interaction.

A result of General Relativity is that the gravitational field is unlocalized [3]. Unlocalized in the Universe means that the Compton wavelength of the quantum of gravity of mass $m_g$, $\hbar/m_g c$, [4], must be of the order of magnitude of the size of the Universe, ct, a constant in our approach. We will prove that this quantum condition is present at the initial stages of the Universe. Then the quantum of gravity, having always this size, does not "collapse". We have proved elsewhere that Zel'dovich and Weinberg's relations are one and the same thing, [5]. They are the result of Newton's laws of motion and gravitation. Now we will prove here that any particle of mass m, obeying either Zel'dovich or Weinberg's relation, has a wave function at the initial stages of the Universe with a width filling all space. They are initially unlocalized, but certainly localized inside the Universe. We will then prove that all particles obeying these relations, except gravity quanta, will collapse due to their Newtonian self-gravitation. The collapse brings the particle to a stable equilibrium stage: the inward acceleration due to self gravitation is balanced by the outward spreading of the wave function. The equilibrium state corresponds to a constant particle size given by its Compton wavelength.

## 2. The first finding: Zel'dovich expression for the cosmological constant Λ

Zel'dovich [6] expressed the cosmological constant Λ in terms of G, m and ℏ, the gravitational constant, the mass of a fundamental particle and Planck's constant, respectively, as follows:

$$\Lambda \approx \frac{G^2 m^6}{\hbar^4} \quad (1)$$

We are taking Planck's constant as a true universal constant. We are lead to this conclusion following the evidence of radioactive materials. There decay time is inversely proportional to the 7$^{th}$ power of ℏ. The observed constancy of this time strongly imposes the constancy of ℏ. On the other hand the product $Gm^3$ is a theoretical universal constant. Otherwise we could not derive Einstein's field equations from the action principle. And lunar laser ranging experiments imply this result too. The constancy of momentum, mc, implies that $G/c^3$ is a universal constant. From (1) the cosmological constant Λ is also a true universal constant. Any cosmological time variation in G implies a time variation in c, to keep $G/c^3$ as a universal constant. And this implies a time variation in m. Whatever these time variations may be, from (1) we see that the cosmological constant is a true universal constant in any case.

It is well known that the value of Λ is of the order of $10^{-56}$ cm$^{-2}$. From the Einstein cosmological equations one has that Λ is of the order of $1/(ct)^2$. This is a property widely used in many works. Another way to look at it is to consider the relation $\Lambda c^2 \approx H^2$. It implies that the product ct is a universal constant, and therefore c varies as 1/t. This is a result we have found elsewhere, [5] and [2]. It gives a linear increase in m, the mass-boom. Hence the size of the Universe given by ct is a constant, of the order of its gravitational radius. It is clear that the speed of light c, being inversely proportional to cosmological time t, must be proportional to the Hubble parameter H, as proved elsewhere, [2]. This is also the conclusion one arrives at when considering the well known relation $\Lambda c^2 \approx H^2$. And if we eliminate the mass m using Zel'dovich and Weinberg's relation, this is exactly the result one gets.

The standard laboratory considerations, the individual constancies of G, m and ℏ, give also from (1) a constant value for Λ. Our time varying findings for these parameters do not change this result. In fact they are completely in line with the Einstein's field equations because they conserve the action principle and the relativity parameter v/c.

## 3. The second finding: Weinberg's relation for the mass of a fundamental particle

Weinberg's relation, [7], is given by

$$m^3 = A\frac{\hbar^2 H}{Gc} \qquad (2)$$

where A is a numerical constant close to one. Comparing this with (1) it is evident that c must be proportional to H. Here we have an important consideration. It is well known that the most simple combination of the parameters G, c and ℏ giving a mass results in the value of Planck's mass $(\hbar c/G)^{1/2}$, which is about 20 orders of magnitude higher than the mass of a fundamental particle m. The presence of Λ in (1) implies that this cosmological constant is a true universal constant. The presence of H, a cosmological parameter, in (2) makes the trick of obtaining the mass m. Since Gm$^3$ is a universal constant, and so is ℏ, the conclusion is that c is also a cosmological parameter like H. In other words, if one is considering the local value of H then one has to consider the local value of c, and vice versa. It is obvious that both parameters are seen as constants at the present age of the Universe. The usual local constancy taken for c is equivalent to take H as a constant too. But they are both cosmologically time varying as 1/t, and there is complete consistency with relativity.

The important point here is that (1) and (2) are the same. And they are the result of Newton's laws of motion and gravitation, as proved elsewhere [2]. In the past it has been very surprising that in the relation (2) there appeared to be only one cosmological parameter, just H, making possible to predict the local value of a typical mass m. We see now that there is another cosmological parameter, c.

## 4. The third finding: mass for balancing inward and outward accelerations

The third finding comes from the computation of a critical mass. It is the critical mass that balances the outward quantum mechanical spreading of the wave function, and its inward gravitational collapse. Carlip [8] has estimated this value. We will identify here this mass with the mass m used by Zel'dovich and Weinberg in (1) and (2).

Carlip [8] starts with the model of semi classical gravity proposed by Møller [9]. The right hand side of the Einstein field equations is replaced by an expectation value:

$$G^{\mu\nu} = 8\pi \langle \psi | \frac{G}{c^4} T^{\mu\nu} | \psi \rangle \qquad (3)$$

Starting with the Newtonian approximation to (3), the Schrödinger-Newton equation given in [10] and [11], we have

$$i\hbar \frac{\partial \psi}{\partial t} = -\frac{\hbar^2}{2m} \nabla^2 \psi - m\Phi\psi, \quad \nabla^2 \Phi = 4\pi Gm|\psi|^2 \qquad (4)$$

This model treats matter quantum mechanically and describes gravity in terms of a classical Newtonian potential Φ. Its source is the expectation value of the mass density.

Considering now the particle m with a localized initial Gaussian wave function, [8], as

$$\psi(r,0) = \left(\frac{\alpha}{\pi}\right)^{3/4} e^{-\alpha r^2/2} \qquad (5)$$

where the width is given by $\alpha^{-1/2}$. We will later prove that $\alpha$ is just the cosmological constant $\Lambda$. The peak probability density for a free particle is given to occur [8] at

$$r_p \approx \alpha^{-1/2}\left(1 + \frac{\alpha^2 \hbar^2}{m^2} t^2\right)^{1/2} \qquad (6)$$

From here we see that there is a time $\tau$ such that for $t < \tau$ the peak probability density occurs at

$$t < \tau$$
$$r_p \approx \alpha^{-1/2} \qquad (7)$$

and for $t \gg \tau$ the peak probability density occurs at

$$t \gg \tau$$
$$r_p \approx \alpha^{1/2} \frac{\hbar}{m} t = \frac{ct}{\alpha^{-1/2}} \frac{\hbar}{mc} \qquad (8)$$

where the dividing reference time $\tau$ is given from (6) by

$$\tau = \frac{m}{\alpha \hbar} \qquad (9)$$

The value of the mass m (at t = 0) that balances the outward acceleration of (6) with the inward gravitational acceleration is given [8] by

$$m \approx \left( \frac{\hbar^2 \sqrt{\alpha}}{G} \right)^{1/3} \qquad (10)$$

We identify this expression with the Zel'dovich (1) and Weinberg (2) relations, with the result of $\alpha \approx \Lambda \approx 1/(ct)^2$.
This is an important step: we identify this mass found by Carlip, [8], with the Weinberg's mass in (2) to obtain

$$\alpha \approx \left( \frac{H}{c} \right)^2 \approx \Lambda, \quad \alpha^{-1/2} \approx \frac{c}{H} \approx \frac{1}{\Lambda^{1/2}} \approx ct \qquad (11)$$

The most important conclusion from this step is that the width of the free particle wave function, as seen above, is always of the order of ct, the size of the Universe. We note that this width is equal to the peak probability density given in (7) for $t < \tau$ and at this age the obvious conclusion is that the free particle fills the whole Universe. On the other hand, for $t \gg \tau$ we get from (8) that the peak probability density is precisely the Compton wave length $\hbar/mc$. This is the present state of the fundamental particles in our Universe

The value of the dividing time $\tau$ between the two conditions (7) and (8) is then

$$\tau = \frac{m}{\alpha \hbar} \approx (ct)^2 \frac{m}{\hbar} = \frac{ct}{\hbar/mc} t \qquad (12)$$

The second important step is now to consider that the time t in (6) is the cosmological time. Then, for $t = \tau$ one gets from (12) that the dividing age of the Universe, that we have defined as $\tau$, occurs when

$$ct = \frac{\hbar}{mc} \qquad (13)$$

This means that the Compton wavelength of the mass m is of the order of the size of the Universe. This is the case given in (7): for $t \leq \tau$ the particle m fills the whole Universe. We also see that relation (13) coincides with the expression of the mass $m_g$ for the quantum of gravity, [4], $m_g = \hbar/c^2 t$.

We will now prove that the time τ is just the Planck's time at this age of the Universe, t = τ.

The relevance of this important result to cosmology can not be overestimated. It says that the peak probability density for a free particle, at the initial stages of the Universe, is equal to the width of its wave function, and equal to the initial size of the Universe. Initially it fills the whole Universe. And this initial size must be a fluctuation: the constant Planck's length given by $(G\hbar/c^3)^{1/2} \approx 10^{-33}$ cm. This is the initial width $r_p$. The age of the Universe at this stage must be the initial value of Planck's time $t_1$:

$$t_1 = \left(\frac{G\hbar}{c^5}\right)^{1/2} \approx 10^{-44} \frac{t}{t_0} \sec \qquad (14)$$

Here $t_0$ is the present age of the Universe. In terms of the present tic-tac interval for fundamental particles it has a value of $10^{41}$ oscillations. Hence the value of the initial Planck's interval, from (14) is

$$t_1 \approx 10^{-105} \sec \qquad (15)$$

This is the length of the first time oscillation during which a fluctuation of size $10^{-33}$ occurred. To arrive at the present size of the Universe, about $10^{28}$ cm, we have to think of an initial inflation by a factor of about $10^{61}$. According to the theory of inflation, this must have occurred during a few tens of time units given by (15). During this interval the width $r_p$ given by (6) jumped from $10^{-33}$ cm to $10^{28}$ cm.

From (7) and (11) we have

$$\begin{array}{rcl} t & < & \tau \\ r_p & \approx & ct \end{array} \qquad (16)$$

and from (8) and (11)

$$\begin{array}{rcl} t & \gg & \tau \\ r_p & \approx & \dfrac{\hbar}{m_p c} \end{array} \qquad (17)$$

From (16) and (17) it is seen that the wave function collapses from ct to ℏ/mc. After inflation the mass m occupies a constant size given by its Compton wave length. The conclusion is that any particle of mass m, obeying the Zel'dovich-Weinberg relations, occupy at the initial stages of the Universe all space. As time goes inflation occurs in a few tens of units of time given in (15), and then it collapses to a size given by its Compton wave length ℏ/mc. This is the history of the present fundamental particles in the Universe that can be traced back to its initial stages: an initial inflation of a Planck's fluctuation followed by a gravitational collapse, to arrive at the size given by its Compton wave length. Our critical time τ is just the initial Planck's unit of time (15).

## 5. Conclusion

The three findings presented here converge: Newton's laws of motion and gravitation, while explaining Zel'dovich and Weinberg's relations, [2]. They also define the mass and size of the fundamental particles when using the Schrödinger-Newton equation given in [10] and [11]. The fundamental particles obey the Zel'dovich and Weinberg's relations. They are born at the initial stages of the Universe as a Planck fluctuation that inflates to the present size of the Universe. Gravitational collapse follows bringing the fundamental particles to their present size. These findings should help to model a quantum-gravitational theory.